\RequirePackage[2020-02-02]{latexrelease}
\documentclass[aip,apl,amsmath,amssymb,twocolumn,reprint]{revtex4-2}

\usepackage{graphicx}
\usepackage{dcolumn}
\usepackage{bm}
\usepackage[utf8]{inputenc}
\usepackage[T1]{fontenc}
\usepackage{mathptmx}
\usepackage{etoolbox}

\makeatletter
\def\@email#1#2{%
 \endgroup
 \patchcmd{\titleblock@produce}
  {\frontmatter@RRAPformat}
  {\frontmatter@RRAPformat{\produce@RRAP{*#1\href{mailto:#2}{#2}}}\frontmatter@RRAPformat}
  {}{}
}%
\makeatother

\preprint{AIP/123-QED}

\begin{document}

%\preprint{}

\title{Radiative coupling the easy way: Using transfer coefficients to model series-connected multi-junction solar cells} %Title of paper



\author{Rune Strandberg}
\email{runes@uia.no}
%\thanks{}
\affiliation{University of Agder, Department of Engineering Sciences, Jon Lilletuns vei 9, NO-4879 Grimstad, Norway }

\date{\today}

\begin{abstract}
When the quality of multijunction solar cells becomes sufficiently high, radiative exchange of photons between cells has to be taken into account to properly model these devices. In this work it is shown how this radiative coupling can be accounted for in series connected multi-junction solar cells by constants called transfer coefficients. Under the assumption that the exchanged radiation only travels one way, from higher to lower band gaps, the transfer coefficients allows the relation between the voltage and current of the device to be expressed by a convenient mathematical expression. Another advantage of this model is that it allows the short circuit current of radiatively coupled multijunction cells to be calculated in a straightforward way. Non-ideality may be included by means of the external radiative efficiency. A number of examples are discussed to show how the transfer coefficients describes the radiative coupling. In addition, the model is used to find the efficiency peaks of radiatively coupled multi-junction cells when illuminated by the AM1.5 spectrum.

\end{abstract}


\maketitle

%\section{Introduction}

In this letter, a model that describes the behaviour of series-connected multijunction solar cells is derived. The approach is inspired by the model previously published by Friedman et al. in Ref. \cite{Friedman2013}, where an inverse $JV$-characteristic is applied. The starting point of the work in Ref. \cite{Friedman2013} is the double diode equation, whereas the present work is based on the single diode model. The latter allows the impact of the radiative coupling to be captured by a few parameters with an interpretation that will be discussed later. 

Following the traditional single diode model, the individual cells in a series connected multi-junction device have current densities $J_i$ that follows 
\begin{equation}
\label{eq:diode}
J_i=J_{\mathrm{G},i}-J_{0,i} e^{\frac{qV_i}{kT}},
\end{equation}
when they are operated as single junction cells \cite{shockley1949,shockley1961}. $J_{\mathrm{G},i}$ is the generation current density experienced by the cell, $J_{0,i}$ its recombination current density in equilibrium \cite{Cuevas2014} and $V_\mathrm{i}$ is the cell voltage. 

When more than one cell are stacked into a series-connected multi-junction device, the device voltage $V$ is the sum of the voltages from the individual cells. Inverting the $JV$-characteristic in Eq. (\ref{eq:diode}) gives the $VJ$-characteristic
\begin{equation}
\label{eq:singleVJ}
V_i=\frac{kT}{q}\mathrm{ln}\left[\frac{J_{\mathrm{G,i}}-J_i}{J_0}\right].
\end{equation}
Assuming that radiative coupling does not occur, the $VJ$-characteristic of a multijunction stack is then found by
\begin{equation}
\label{eq:VJsimple}
V=\sum_i V_i=\frac{kT}{q}\mathrm{ln}\left[\prod_i \frac{J_{\mathrm{G,i}}-J}{J_0}\right],
\end{equation}
an expression previously used by Fernandez et al. \cite{Fernandez2013}, which also represents a special case of the model in Ref. \cite{Friedman2013}.
%As long as the individual cells obey Eq. (\ref{eq:diode}), Eq. (\ref{eq:VJsimple}) can describe a stack with an unlimited number of cells.
If the stack contains two cells, Eq. (\ref{eq:VJsimple}) can be solved with respect to $J$ to give the $JV$-characteristic 
\begin{equation}
\label{eq:IVtandem}
J=\frac{1}{2}\left(J_\mathrm{G,1}+J_\mathrm{G,2}\right)-\sqrt{\frac{1}{4}\Delta J_\mathrm{G}^2+J_{0,1}J_{0,2}\mathrm{e}^{\frac{qV}{kT}}},
\end{equation}
where $\Delta J_\mathrm{G}=J_\mathrm{G,1}-J_\mathrm{G,2}$, as in Ref. \cite{Strandberg2020}. For stacks with more cells, finding the $JV$-characteristic requires higher order equations to be solved, which leaves keeping $VJ$-characteristic as in Eq. (\ref{eq:VJsimple}) a more convenient option. 
If radiative coupling is present, the $JV$-characteristic of a top cell in an ideal device can be expressed as \cite{Strandberg2020}
\begin{equation}
\label{eq:JVRCtop}
J_1=J_{\mathrm{G},1}-\left(1+n^2\right)J_{0,1}e^{\frac{qV_1}{kT}} + n^2J_{0,1}e^{\frac{qV_2}{kT}},
\end{equation}
where $n$ is the refractive index. Assuming $n$ to be constant throughout the stack, the bottom cell in a stack of $N$ cells is described by
\begin{eqnarray}
\label{eq:JVRCN}
J_N &=& J_{\mathrm{G},N}-\left(J_{0,N}+n^2J_{0,N-1}\right)e^{\frac{qV_N}{kT}} \nonumber \\
&&+ n^2J_{0,N-1}e^{\frac{qV_{N-1}}{kT}},
\end{eqnarray}
while the remaining cells follow
\begin{eqnarray}
\label{eq:JVRCi}
J_i&=&J_{\mathrm{G},i}-\left[\left(1+n^2\right)J_{0,i}+2n^2J_{0,i-1}\right]e^{\frac{qV_i}{kT}} \nonumber \\
&&+ n^2J_{0,i-1}e^{\frac{qV_{i-1}}{kT}} + n^2J_{0,i}e^{\frac{qV_{i+1}}{kT}}.
\end{eqnarray}
(For Eqs. (\ref{eq:JVRCN}) and (\ref{eq:JVRCi}) to be accurate, the integration limits used when calculating the radiative component of $J_{0,i}$ has to be set to the band gaps of cell $i$ and cell $i-1$. Details are found in Ref. \cite{Strandberg2020}.)
Since the series-connection dictates that $J_1=J_i=J_N := J$, Eqs. (\ref{eq:JVRCtop})-(\ref{eq:JVRCi}) constitutes a set of equations which are linear in the $N$ unknowns $\mathrm{exp}\left( qV_i/kT \right)$. From the solutions of this set, the $VJ$-characteristic can be constructed by 
\begin{equation}
\label{eq:VJgeneral}
V= \frac{kT}{q}\mathrm{ln}\left(\prod_i e^\frac{qV_i}{kT} \right)
\end{equation}
where $V_i$ are functions of $J$. As above, reorganizing the solution for a stack of two cells to isolate $J$, gives the $JV$-characteristic derived in Ref. \cite{Strandberg2020} for devices with radiative coupling. For stacks with more cells, the set of equations is easily solved numerically. To further explore the behavior of radiatively coupled cells analytically, it is convenient to exploit the fact that in useful devices, practically all the luminescence in the stack propagates from cells with higher band gaps to neighboring cells with smaller band gaps, i.e. from cell $i$ to cell $i+1$. This is true as long as $V_i$ is more than a few times $kT$ larger than $V_{i-1}$. Since the point of a multi-junction device is to extract the current generated by high energy photons at a higher voltage, this criterion is normally fulfilled for meaningful band gap combinations. Beware, however, that exceptions may exist, for example if the stack has a large number of cells.

When the radiative coupling works in only one direction, we can solve for $\mathrm{exp}\left( qV_i/kT \right)$ one cell at the time, starting at the top. To generalize the approach, the cells in the stack will be allowed have different refractive indices. $n_i$ is used for the refractive index which limits the transfer of photons to cell $i$. From the system of equations, the first unknown is found from (\ref{eq:JVRCtop}) as
\begin{equation}
\label{eq:simplsolX1}
e^{\frac{qV_1}{kT}}=\frac{J_\mathrm{G,1}-J}{\left(1+n_2^2\right)J_{0,1}}.
\end{equation}
For the next cell, assuming it is not the bottom cell, one arrives at
\begin{equation}
\label{eq:simplsolX2}
e^{\frac{qV_2}{kT}}=\frac{J_\mathrm{G,2}-J+\frac{n_2^2}{n_2^2+1}\left(J_\mathrm{G,1}-J \right)}{\left(1+n_3^2\right)J_{0,2}}.
%X_2=\frac{J_\mathrm{G,2}-J-\mathcal{T}\Delta J_\mathrm{G,2}}{\left(1+n^2\right)\left(1-\mathcal{T}\right)J_{0,2}}
\end{equation}
The fraction of the surplus photons from the top cell being transferred to the next cell is equal to $n_2^2/(n_2^2+1)$. Eq. (\ref{eq:simplsolX2}) can be rewritten by introducing a transfer coefficient $T_2$ that equals the fraction of the current mismatch $\Delta J_{\mathrm{G_2}}=J_{\mathrm{G_1}}-J_{\mathrm{G_2}}$ which is transferred to the next cell. The result is 
\begin{equation}
\label{eq:reorgsol2}
e^{\frac{qV_2}{kT}}=\frac{J_\mathrm{G,2}+T_2\Delta J_\mathrm{G,2}-J}{\left(1-T_2\right)\left(1+n_3^2\right)J_{0,2}},
\end{equation}
where
\begin{equation}
\label{eq:tranfercoeff}
T_2=\frac{n_2^2}{1+2n_2^2}:=\mathcal{T}_2.
\end{equation}
The reason for defining this fraction as both $T_2$ and $\mathcal{T}_2$ will become clear below.
The numerator in Eq. \ref{eq:reorgsol2} is now written as the difference between the total generation current experienced by cell 2 and the current flowing through the device, analogous to Eqs. (\ref{eq:VJsimple}) and (\ref{eq:simplsolX1}). For cells further down the stack the picture is a bit more complicated because the impact of the radiative transfer propagates downwards. Some ink can be saved by defining $\tilde{J}_{0,i}=(1+n_{i+1}^2)^{1-\delta(N-i)}J_{0,i}$. The Dirac delta function $\delta(N-i)$ appears here since the bottom cell is not emitting luminescence to a cell further down the stack. Instead it is assumed to have a perfect reflector at its back side.

Proceeding to the solution for the third cell now gives 
\begin{equation}
\label{eq:simplsolX3}
e^{\frac{qV_3}{kT}}=\frac{J_\mathrm{G,3}+T_3\Delta J_\mathrm{G,3}-J}{\left(1-T_3\right)\tilde{J}_{0,3}},
\end{equation}
with
\begin{equation}
\label{eq:K3}
T_3=\frac{\mathcal{T}_3}{1+(\mathcal{T}_3-1)T_2}
\end{equation}
and
\begin{equation} 
\label{eq:delta3}
\Delta J_\mathrm{G,3}=J_\mathrm{G,2}+T_2\Delta J_\mathrm{G,2}-J_\mathrm{G,3}.
\end{equation}
The value of $\mathcal{T}_3$ is given by a straightforward generalisation of Eq. (\ref{eq:tranfercoeff}), i.e.
\begin{equation}
\label{eq:tranfercoeffgeneral}
\mathcal{T}_i=\frac{n_i^2}{1+2n_i^2}.
\end{equation}
The larger the value of $\mathcal{T}_i$, the stronger the radiative coupling into cell $i$.

Cells further down the stack of $N$ cells follow the same pattern, which allows the generalization  
\begin{equation}
\label{eq:Xi}
e^{\frac{qV_i}{kT}}=\frac{J_\mathrm{G,i}+T_i\Delta J_{\mathrm{G},i}-J}{\left(1-T_i\right)\tilde{J}_{0,i}}
\end{equation}
with
\begin{equation}
\label{eq:defK}
T_i=\frac{\mathcal{T}_i}{1+(\mathcal{T}_i-1)T_{i-1}},
\end{equation}
where the latter is a recursive formula initiated by $T_1=0$. $\Delta J_{\mathrm{G},i}$ is the difference between the total generation current density in cell $i-1$ and the external generation current density in cell $i$, and thus given by
\begin{equation}
\label{eq:deltai}
\Delta J_{\mathrm{G},i}=J_{\mathrm{G},i-1}+T_{i-1}\Delta J_{\mathrm{G},i-1}-J_{\mathrm{G},i}
\end{equation}
for $i>1$, whereas $\Delta J_\mathrm{G,1}:= 0$.

The $VJ$-characteristic of the entire stack can now be conveniently written as
\begin{equation}
\label{eq:VJRCapprox}
V=\frac{kT}{q}\mathrm{ln}\left[\prod_{i=1}^N\frac{J_\mathrm{G,i}+T_i\Delta J_{\mathrm{G},i}-J}{\left(1-T_i \right)\tilde{J}_{0,i}}\right].
\end{equation}
For a perfectly current-matched device where all the cells have the same generation current $J_\mathrm{G}$, Eq. (\ref{eq:VJRCapprox}) can be solved for $J$, which yields the $JV$-characteristic
\begin{equation}
\label{eq:JVCM}
J=J_\mathrm{G}-\left(\prod_{i=1}^N \left(1-T_i \right)\tilde{J}_{0,i}\right)^{\frac{1}{N}}e^{\frac{qV}{NkT}},
\end{equation}
which is on the form of the $JV$-characteristic of a single-junction solar cell with an ideality factor of $N$. Consequently, expressions for the maximum power point on the form found by Khanna et al. \cite{Khanna2013} for single junctions cells, can be applied for this special case, as can textbook expressions for the open circuit voltage. For stacks that are not current matched, the open circuit voltage is easily found by setting $J=0$. The short circuit current density is found when $V=0$, which implies that the argument of the logarithm must equal 1. Since the values of $J_{\mathrm{G},i}$ are several orders of magnitude larger than the values of $J_{0,i}$, the numerator of the factor belonging to the current-limiting cell in Eq. (\ref{eq:VJRCapprox}) must be so close to zero at short circuit that this small difference is negligible. It follows that the short circuit current is given by
\begin{equation}
\label{eq:Jsc}
J_\mathrm{sc}=\mathrm{min}\{ J_\mathrm{G,i}+T_i \Delta J_{\mathrm{G},i} \}.
\end{equation} 
\begin{figure}[htbp]
\includegraphics[width=9cm]{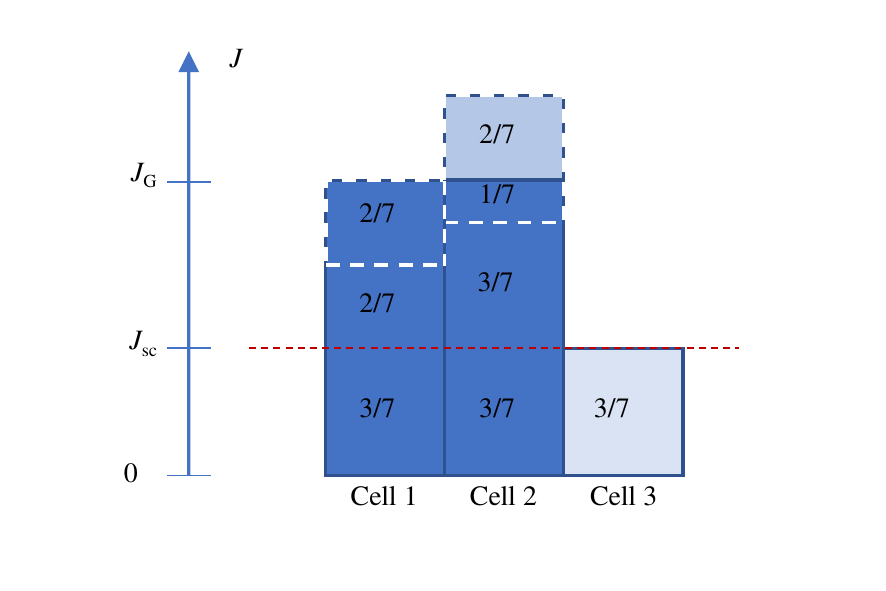}
\caption{In this example, cell 1 and cell 2 have an external generation current $J_\mathrm{G}$, while cell 3 receives no photons from the external light source. A refractive index of 1 is assumed for simplicity. Half of the photon surplus absorbed by cell 1 is transferred to cell 2, as marked by the dashed rectangle on the bar of cell 1. The other half is emitted to the surroundings. This gives a total generation in cell 2 amounting $\tfrac{9}{7}J_\mathrm{G}$, where the 2/7 coming from the top cell is marked by a lighter color. Cell 2 also emits half of its surplus, the 3/7 marked by the dashed rectangle, to the cell below. The current-limiting cell 3 thus ends up with a short circuit current equaling $\tfrac{3}{7} J_\mathrm{G}$. In total, photons equivalent to $\tfrac{9}{7}J_\mathrm{G}$ are kept within the cell, while a flux equivalent to $\tfrac{5}{7}J_\mathrm{G}$ is emitted to the surroundings.  \label{fig:transfer}}
\end{figure}
\begin{table}
\caption{Parameter values to guide the interpretation of Fig. \ref{fig:transfer2}. The refractive index is set to 3 in this example. \label{tab:JSc2}}
\begin{ruledtabular}
\begin{tabular}{ccccrc}
Cell number& $J_{\mathrm{G},i}$ & $\mathcal{T}_i$ &  $T_i$ & $\Delta J_{\mathrm{G},i}$ & $J_\mathrm{G,i}+T_i\Delta J_{\mathrm{G},i}$ \\ 
\hline \\[-4pt]
1& $2J_{\mathrm{G}}$ & 0 & 0 & 0 & $2J_{\mathrm{G}}$  \\[4pt]
2& $J_{\mathrm{G}}$  & $\frac{9}{19}$ & $\frac{9}{19}$ & $J_{\mathrm{G}}$ & $\tfrac{28}{19}J_{\mathrm{G}}$ \\[4pt]
3& $J_{\mathrm{G}}$  & $\frac{9}{19}$ & $\frac{171}{271}$ & $\tfrac{9}{19}J_{\mathrm{G}}$ & $\tfrac{352}{271}J_{\mathrm{G}}$  \\[4pt]
\end{tabular}
\end{ruledtabular}
\end{table}
% \begin{figure}[htbp]
% \includegraphics[width=9cm]{Transfer2}
% \caption{In this example the top cell receives a photon flux equivalent to a generation current $J_\mathrm{G,1}$ from an external light source. This is twice as large as the external generation current $J_\mathrm{G,2}$ experienced by cell 2. Cell 3 receives no external radiation. The refractive index is set to 1. Cells 1 and 2 will emit half of their surplus photons towards the cell below and the other half to the surroundings. In fractions of $J_\mathrm{G,2}$, cell 1 emits $\tfrac{5}{7}$ to cell 2, which gives cell 2 a surplus of $\tfrac{8}{7}$, of which $\tfrac{4}{7}$ is transferred to cell 3. The short current thus becomes $\tfrac{4}{7}J_\mathrm{G,2}$ \label{fig:transfer2}}
% \end{figure}
\begin{figure}[htbp]
\includegraphics[width=9cm]{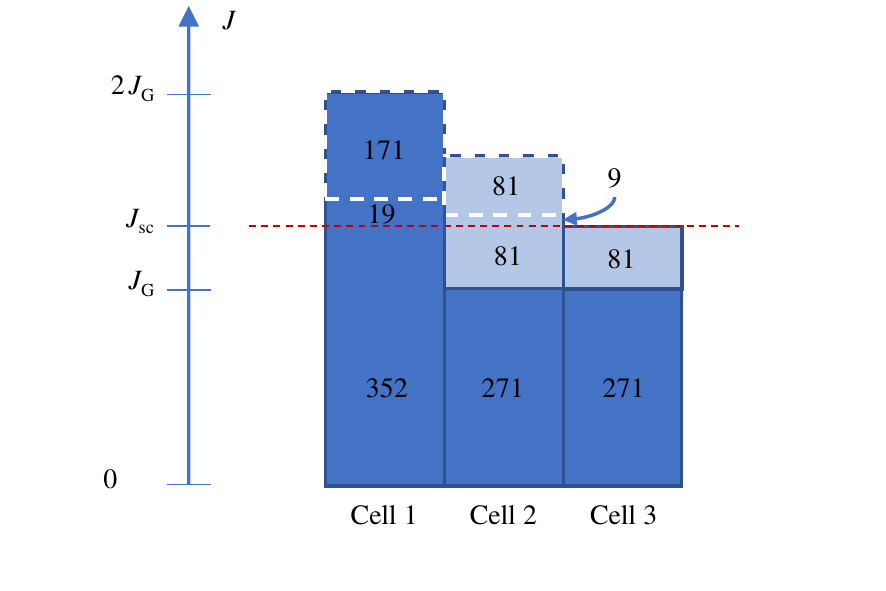}
\caption{In this example the middle and bottom cells receive a photon flux equivalent to a generation current density of $J_\mathrm{G}=271\,\mathrm{mA/cm^2}$ from an external light source. The generation current of the top cell is twice as large. The refractive index is set to 3 in this example. After computing the values in Tab. \ref{tab:JSc2}, the short circuit current-density is found and the radiative transfer between the cells can be backtracked. The numeric values in the figure show the equivalent current densities, in $\mathrm{mA/cm^2}$, involved in different processes. The dashed rectangles indicates contributions emitted to the cell below, whereas the lighter color indicates contributions received from the cell above. The fields between the line indicating $J_\mathrm{sc}$ and the dashed rectangles corresponds to the photons emitted to the surroundings at short-circuit. \label{fig:transfer2}}
\end{figure}
The transfer coefficient $T_i$ can thus be interpreted as the fraction of $\Delta J_{\mathrm{G},i}$ that is transferred to cell $i$ at short circuit - provided that cell $i$ is the current limiting cell. Another interpretation can be found if one assumes that the cells above cell $i$, that is cells 1 to $i-1$, are current-matched. $\Delta J_{\mathrm{G},j}$ is then zero for $j<i$. If cell $i$ has a generation current $J_{\mathrm{G,i}}<J_{\mathrm{G,j}}$, $T_i$ will equal the fraction of the difference $J_{\mathrm{G,j}} - J_{\mathrm{G,i}}$ that is transferred to cell i as a result of the luminescent coupling. An illustration with a triple junction device is shown in figure \ref{fig:transfer}. Here the two upper cells are current-matched, whereas the third cell receives no illumination at all from external sources. A refractive index of 1 is assumed for simplicity. This gives the transfer coefficients $T_2=1/3$ and $T_3=3/7$. The top cell will emit one half of its surplus photons to the surroundings and the other half to the next cell, in accordance with the observation made after Eq. (\ref{eq:simplsolX2}). After receiving this additional photon flux, the middle cell have a larger surplus of photons than the top cell. This larger surplus is also split in two, and the bottom cell receives half of it. As seen from the figure, the short circuit current of this device becomes 3/7 of the generation current of the upper two cells, as it should be according to the value of $T_3$. 

Another example is shown in figure \ref{fig:transfer2}. Here the middle and bottom cells receive an external generation current density $J_\mathrm{G}$, while the top cell receives a generation current density that is twice as large. Assuming a refractive index of 3 allows the values in Tab. \ref{tab:JSc2} to be calculated. From the table it becomes clear that the bottom cell is the bottleneck of this device. If we assume that $J_\mathrm{G}=271\,\mathrm{mA/cm^2}$, the short circuit current density becomes $352\, \mathrm{mA/cm^2}$. This means that photons corresponding to $81\,\mathrm{mA/cm^2}$ has been radiatively transferred from cell 2 to cell 3. With a refractive index of 3, this transfer should correspond to 9/10 of the photons emitted from the middle cell, meaning photons corresponding to $9\,\mathrm{mA/cm^2}$ has been emitted from this cell to the surroundings. To assure current-matching, additional photons equivalent to $81\,\mathrm{mA/cm^2}$ need to be transferred to the cell in the middle. A total of $171\,\mathrm{mA/cm^2}$ must therefore have been transferred from the top cell to the middle cell. Also the top cell should emit 9/10 of its total emission towards the cell below, which implies that an equivalent of $19\,\mathrm{mA/cm^2}$ is emitted to the surroundings. This leaves $352\,\mathrm{mA/cm^2}$ in the top cell - exactly what it needs to be current-matched to the other cells.

\begin{table}
\caption{Parameter values to guide the interpretation of Fig. \ref{fig:transfer3}. The refractive index is set to 1. \label{tab:JSc3}}
\begin{ruledtabular}
\begin{tabular}{ccccrc}
Cell number& $J_{\mathrm{G},i}$ & $\mathcal{T}_i$ &  $T_i$ & $\Delta J_{\mathrm{G},i}$ & $J_\mathrm{G,i}+T_i\Delta J_{\mathrm{G},i}$ \\ 
\hline \\[-4pt]
1& $J_{\mathrm{G}}$ & 0 & 0 & 0 & $J_{\mathrm{G}}$  \\[4pt]
2& 0 & $\frac{1}{3}$ & $\frac{1}{3}$ & $J_{\mathrm{G}}$ & $\tfrac{1}{3}J_{\mathrm{G}}$ \\[4pt]
3& $J_{\mathrm{G}}$ & $\frac{1}{3}$ & $\frac{3}{7}$ & $-\tfrac{2}{3}J_{\mathrm{G}}$ & $\tfrac{5}{7}J_{\mathrm{G}}$  \\[4pt]
4& 0 & $\frac{1}{3}$ & $\frac{7}{15}$ & $\tfrac{5}{7}J_{\mathrm{G}}$ & $\tfrac{1}{3}J_{\mathrm{G}}$   \\[4pt]
\end{tabular}
\end{ruledtabular}
\end{table}

\begin{figure}[htbp]
\includegraphics[width=9cm]{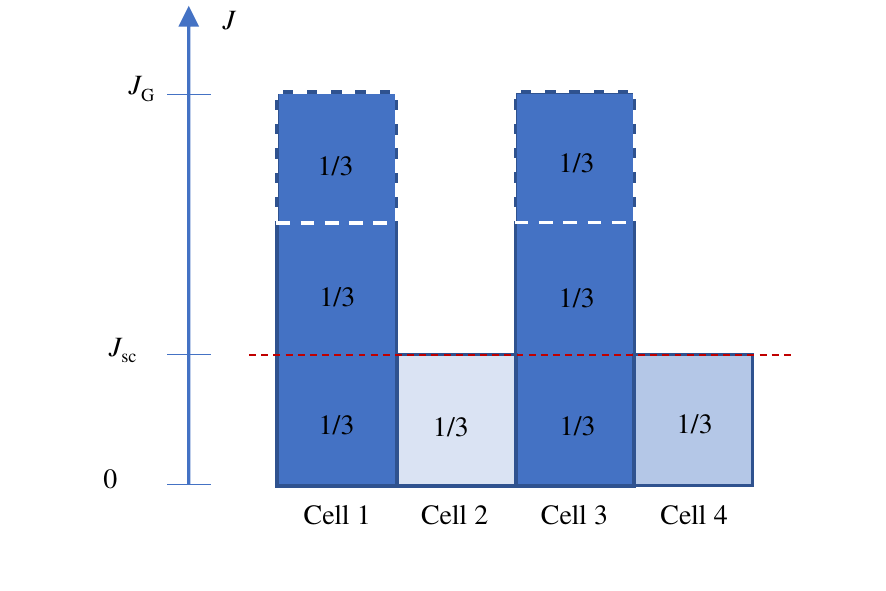}
\caption{A four-junction device where only the first and third cell receives external radiation. The refractive index is set to 1. The generation current of these two cells is $J_\mathrm{G}$. Photons corresponding to one third of the generation current is emitted to the cell below while another third is emitted to the surroundings. The short circuit current equals one third of the generation current.\label{fig:transfer3}}
\end{figure}

As a last example, the situation in Fig. \ref{fig:transfer3} will be considered. This is included to see what happens when $\Delta J_{\mathrm{G},i}$ is negative for one of the cells. The refractive index is once again assumed to be 1, as in the first example. In this stack of four cells, cells 1 and 3 receives external radiation corresponding to a generation current $J_\mathrm{G}$, while the other two cells receive no external radiation at all. The short circuit current in this device is $\tfrac{1}{3}J_\mathrm{G}$, and it is limited by both cell 2 and 4, because one half of the excess photons in cells 1 and 3 is transferred to the cell below while the other half is emitted to the surroundings. Relevant parameters are listed in Tab. \ref{tab:JSc3}. In this case the effective generation current ($J_\mathrm{G,3}+T_3\Delta J_{\mathrm{G},3}$) of cell 3 becomes $\tfrac{5}{7}J_\mathrm{G}$. since $T_4 = 7/15$ for this device, the effective generation current of cell four ends up at $T_4 \Delta J_\mathrm{G,4} = 1/3$, as it should be.

Introducing the external radiative efficiency (ERE) enables the inclusion of non-radiative recombination in the modeling. The ERE was originally defined for single junction cells as the fraction of the total number of net recombination events that leads to emission of photons from the cell \cite{Green2012}. For radiatively coupled cells in a multijunction stack, it makes sense to slightly redefine the ERE as the fraction of the net recombination events that lead to emission of photons \emph{to the surroundings}. That is, photons emitted to another cell is not taken into account when calculating the ERE. This assures that a cell of a particular quality will be characterized by the same ERE regardless of whether it is utilized as a single junction cell with a reflector at the back, or as a cell in a multijunction device radiatively coupled to other cells. A similar definition and approach was used in Ref. \cite{Pusch2019}.

The incorporation of the ERE is done by a couple of smaller tweaks. Firstly, non-radiative recombination increases the dark current density, so $\tilde{J}_{0,i}$ must be modified. From the modified definition of the ERE it becomes clear that $\tilde{J}_{0,i}$ is now given by
\begin{equation}
\label{eq:J0ERE}
\tilde{J}_{0,i}=\left(\frac{1}{\mathrm{ERE}_i}+n^2\left[1-\delta(N-i)\right]\right)J_{0,i}
\end{equation}
where $\mathrm{ERE}_i$ is the ERE of cell $i$. Secondly, non-radiative recombination reduces the fraction of surplus photons that may be transfered from one cell to another, which leads to smaller transfer coefficients. Using Eq. (\ref{eq:J0ERE}) when deriving the transfer coefficients gives
\begin{equation}
\label{eq:ERETC}
\mathcal{T}_i=\frac{n_i^2}{\frac{1}{\mathrm{ERE}_i}+2n_i^2}.
\end{equation}
All other expressions are unaffected by the introduction of non-radiative recombination by means of the ERE. Note that a combination of a refractive index of 3 and an ERE equal to 1/9 gives $\mathcal{T}=1/3$. Figs. \ref{fig:transfer} and \ref{fig:transfer3} are therefore valid for this combination of parameters as well. There is a difference in the interpretation of the figures, however, because 8/9 of the recombination that lead to emission of photons to the surroundings from an ideal cell is now replaced by non-radiative recombination. In Fig. \ref{fig:transfer3} for example, 8/27 of $J_\mathrm{G}$ is lost to non-radiative recombination in cells 1 and 3. Photons corresponding to 1/27 of $J_\mathrm{G}$ are emitted to the surroundings from each of these cells, with the new set of parameters.

In principle, the transfer coefficients can be determined experimentally. To determine the value of $T_i$, the external illumination of cell $i$ should be blocked, while all cells above cell $i$ should experience the same external generation current density $J_\mathrm{G}$. The cells below cell $i$ need to absorb enough photons to avoid being a bottleneck. After measuring the short-circuit current, the transfer coefficient can be calculated from Eq. (\ref{eq:Jsc}) as $T_i=J_\mathrm{sc}/J_\mathrm{G}$.

\begin{figure}[htbp]
\includegraphics[width=9cm]{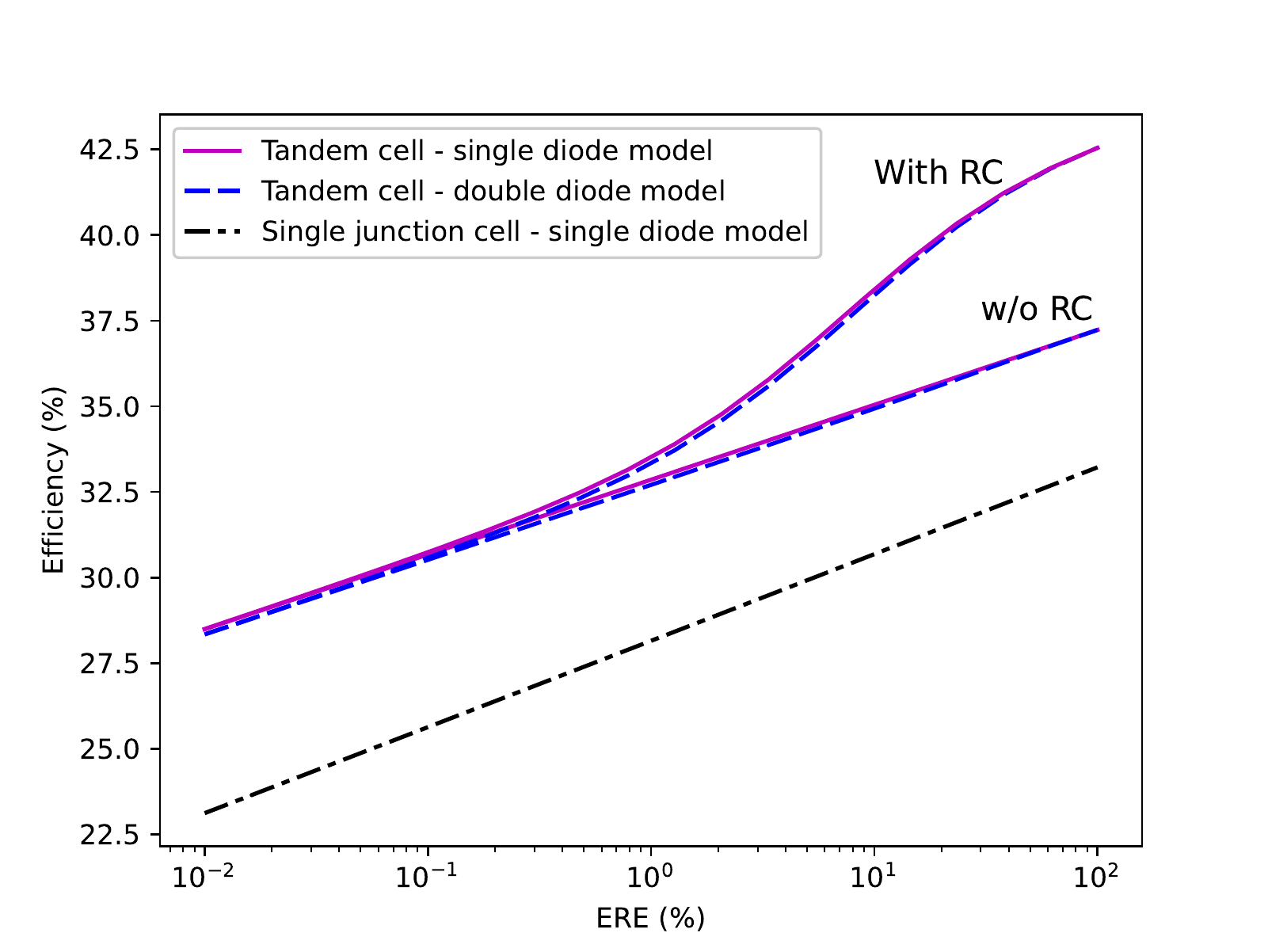}
\caption{The efficiency plotted as a function of the ERE for three different cases. The upper pair of curves are calculated for a device with band gaps of 1.60 and 0.94 eV, which is close to the optimum for double junction devices. The middle pair of curves are calculated with band gaps of 1.60 and 1.11 ev. This is a situation where the generation current densities of the top and bottom cells has a bit of mismatch which is mitigated by radiative coupling. Note that the efficiency of this band gap combination may be improved by thinning the top cell, as described in Ref. \cite{Kurtz1990}, to allow more photons to proceed to the bottom cell. The curve at the bottom is calculated for a single junction device with a band gap of 1.11 eV.   \label{fig:ERE_Eff}}
\end{figure}

The power density delivered by series-connected multi-junction solar cells can be calculated by multiplying Eq. (\ref{eq:VJRCapprox}) by $J$ and optimizing the resulting function. Figure \ref{fig:ERE_Eff} shows a plot of the efficiency as a function of the ERE for tandem cells with band gaps of 1.60 and 1.11 eV. The AM1.5G spectrum has been assumed. The solid curves show the efficiency, for devices with and without radiative coupling, calculated using the model presented in this article. If the ERE is smaller than about $0.1\,\%$, the the radiative coupling becomes irrelevant. Since non-ideal cells may be better described by a two-diode model, a small comparison to such a model is given by the dashed lines in Fig. \ref{fig:ERE_Eff}. These lines are calculated by applying an extreme version of the model from Ref. \cite{Friedman2013}, assuming that all non-radiative recombination is taken care of by the second diode. The ERE of the double diode model is calculated at the maximum power point. The difference between the solid and corresponding dashed lines is smaller than 0.2 percentage points. In the two-diode model, the ERE is voltage dependent and calculating the ERE at the open circuit voltage, instead of at the MPP, increases the difference by about an order of magnitude due to the different shapes of the JV-characteristics of the two models. The good agreement between the models in Fig. \ref{fig:ERE_Eff} indicates that using Eq. (\ref{eq:Jsc}) to determine the short circuit current may also give good results for cells with a two-diode behaviour, if the ERE at the maximum power point is used.

\begin{table*}
\caption{Limiting efficiencies found for stacks of ideal series-connected solar cells without radiative coupling when illuminated by the AM1.5G spectrum. The cell temperature is set to $300\,\mathrm{K}.$\label{tab:RC_eff2}}
\begin{ruledtabular}
\begin{tabular}{cccccccc}
 Efficiency (\%)&$E_\mathrm{g,1}$ (eV) & $E_\mathrm{g,2}$ (eV) & $E_\mathrm{g,3}$ (eV) & $E_\mathrm{g,4}$ (eV) & $E_\mathrm{g,5}$ (eV) & $E_\mathrm{g,6}$ (eV) \\
\hline \\ [-5pt]
45.73& 1.632 & 0.960 & & & &  \\
51.62& 1.900 & 1.367 & 0.933 & & &  \\
55.33& 2.002 & 1.494 & 1.115 & 0.716 & &  \\
57.66& 2.140 & 1.667 & 1.331 & 1.011 & 0.703 &  \\
59.55& 2.238 & 1.787 & 1.469 & 1.194 & 0.958 & 0.692 \\
\end{tabular}
\end{ruledtabular}
\end{table*}

\begin{table*}
\caption{Limiting efficiencies found for radiatively coupled stacks of ideal series-connected solar cells when illuminated by the AM1.5G spectrum. The efficiency is calculated by multiplying Eq. (\ref{eq:VJRCapprox}) by $J$ and optimizing the resulting function. A refractive index of 3.4 is assumed, as in Ref. \cite{Bremner2016}. The cell temperature is set to $300\,\mathrm{K}.$\label{tab:RC_eff}}
\begin{ruledtabular}
\begin{tabular}{cccccccc}
 Efficiency (\%)&$E_\mathrm{g,1}$ (eV) & $E_\mathrm{g,2}$ (eV) & $E_\mathrm{g,3}$ (eV) & $E_\mathrm{g,4}$ (eV) & $E_\mathrm{g,5}$ (eV) & $E_\mathrm{g,6}$ (eV) \\
\hline \\ [-5pt]
44.42& 1.585 & 0.940 & & & &  \\
50.02& 1.877 & 1.345 & 0.933 & & &  \\
53.31& 1.985 & 1.479 & 1.114 & 0.722 & &  \\
55.64& 2.107 & 1.633 & 1.268 & 0.983 & 0.696 &  \\
57.71 & 2.215 & 1.765 & 1.450 & 1.176 & 0.944 & 0.692 \\
\end{tabular}
\end{ruledtabular}
\end{table*}

 The limiting efficiencies for cells without radiative coupling published by Bremner et al.\cite{Bremner2008a}, where a full detailed balance model was used, are reproduced by the model derived in this work. This confirms that the diode equation is an excellent approximation for ideal cells illuminated with the unconcentrated AM1.5 spectrum. Note that this may not be the case, however, if the light is concentrated, or if cells with small band gaps are analyzed. In Ref. \cite{Bremner2008a}, peak efficiencies were identified using a grid search method with a band gap resolution of 0.01 eV. To slightly refine the efficiency limits of multi-junction devices subject to the AM1.5G spectrum, the Nelder-Mead method \cite{Nelder1965} was applied to search for efficiency peaks with the model presented above. This implies that peaks have been searched for in a continuum of band gap combinations, rather than in a discrete grid. Table \ref{tab:RC_eff2} lists the results of this search. While most of the results are minor adjustments of the efficiency peaks from Ref. \cite{Bremner2008a}, one notable exception is found for a stack of two cells where a slightly higher limiting efficiency is found at a new peak. For such a configuration a peak efficiency of $45.73\,\%$ is obtained for band gaps of 1.632 eV and 0.960 eV, while Ref. \cite{Bremner2008a} has a peak of $45.71\,\%$ for band gaps of 1.60 eV and 0.94 eV. (The latter efficiency of $45.71\,\%$ is reproduced for the given band gap combination.)

A search was also performed for the optimal band gaps and corresponding efficiencies for radiatively coupled ideal cells. Setting the refractive index to 3.4, a typical value for semiconductors, throughout the stack, gives the results shown in Tab. \ref{tab:RC_eff}. The calculations are based on Eq. (\ref{eq:VJRCapprox}) and the AM1.5G spectrum is used. The efficiency as a function of the band gaps has several local maxima. The search for the global maximum was carried out by generating 1000 (2000 for six band gaps) random start configurations from which the Nelder-Mead method was applied to identify peaks. The number of start configurations assures that all peaks in the table where identified several times. It is possible, but not likely, that even higher peaks exist. Recalculating the efficiency at the efficiency peaks with Eqs. (\ref{eq:JVRCtop}) to (\ref{eq:JVRCi}) gives the same result. The error introduced by assuming one directional luminescent coupling is thus negligible for the optimal band gaps.

% \begin{equation}
% \label{eq:twodiodeJV}
% J_i=J_{\mathrm{G},i}-J_{0,i} e^{\frac{qV_i}{kT}}-J_{\mathrm{nr},i} e^{\frac{qV_i}{2kT}}
% \end{equation}
% as the starting point. $J_{nr,i}$ is a coefficient describing the magnitude of the non-radiative recombination. Reorganizing this $JV$-characteristic into a $VJ$-characteristic yields
% \begin{equation}
% \label{eq:twodiodeVJ}
% V_i=\frac{2kT}{q}\ln \left( \frac{\sqrt{J_{\mathrm{nr},i}^2+4J_{0,i}\left(J_{\mathrm{G},i}-J\right)}-J_{\mathrm{nr},i}}{2J_{0,i}}\right).
% \end{equation}
% Applying the same steps as for the one-diode model above yields
% \begin{equation}
% \label{eq:twodiodeVJ}
% V_i=\frac{2kT}{q}\ln \left(\prod_{i=1}^N \frac{\sqrt{J_{\mathrm{nr},i}^2+4\tilde{J}_{0,i}\left(J_{\mathrm{G},i}^*-J\right)}-J_{\mathrm{nr},i}}{2\tilde{J}_{0,i}}\right)
% \end{equation}
% for a tandem stack with radiatively coupled cells. The modified generation current density $J_{\mathrm{G}^*,i}$ equals $J_{\mathrm{G},1}$ for the top cell, but is given by
% \begin{eqnarray}
% \label{eq:JGstar}
% J_{\mathrm{G},i}^*=&&J_{\mathrm{G},i}+\frac{n^2}{4\left(n^2+1\right)\tilde{J}_{0,i-1}} \nonumber\\
% &&\times \left(\sqrt{J_{\mathrm{nr},i-1}^2+4\tilde{J}_{0,i-1}\left(J_{\mathrm{G},i-1}^*-J\right)}-J_{\mathrm{nr},i-1}\right)^2 
% \end{eqnarray}
% for the remaining cells. Unfortunately, transfer coefficients cannot be obtained for the two-diode model as for the one-diode model, so in this respect the two-diode model falls short in terms of interpretability.

To summarize, it has been shown that the current-voltage characteristic of radiatively coupled series-connected multi-junction solar cells can be described with a short and handy expression by means of Eq. (\ref{eq:VJRCapprox}). The model reproduces the AM1.5G efficiency of ideal cells found in the literature. Non-radiative recombination can be incorporated by the ERE. Note that the present model does not take into account the impact of quasi-Fermi level gradients, which may give rise to additional transfer of photons between cells \cite{Lan2015}.

%Jgstar[i]=Jg[i]+n**2/(4*(1+n**2)**2*J0[i-1])*(np.sqrt(J02[i-1]**2+4*((1+n**2)*J0[i-1]*(Jgstar[i-1]-J)))-J02[i-1])**2

%A review of modeling of luminescent coupling effect in multi-junction solar cell based on diode equation  Zhezhi Liu, Hui Lv, Yue Hu, Jun Liao, Hang Zhou, Yuehong Su
%Schuster 2021: https://www.sciencedirect.com/science/article/pii/S0960148121015871#bib30 Bruker diodeligningen på hver enkelt celle
%\cite{Gonzalez2011}?
%Jäger: https://onlinelibrary.wiley.com/doi/full/10.1002/solr.202000628
%Tillmann: https://ieeexplore.ieee.org/abstract/document/9518891  (0.1 eV reduction in band gap with LC.
%Xia, ERE for multijunction devices: https://ieeexplore.ieee.org/abstract/document/9300433

\section*{References}
\bibliographystyle{IEEEtran}
\bibliography{bibliography}

\end{document}